\documentclass[doublecol]{epl2}
\usepackage{graphicx}% Include figure files
\usepackage{dcolumn}% Align table columns on decimal point
\usepackage{bm}% bold math
\usepackage{color}
\usepackage{ulem}
\title{An amplitude-frequency study of turbulent scaling intermittency using Empirical Mode Decomposition and
Hilbert Spectral Analysis}
\shorttitle{An amplitude-frequency study of turbulent scaling intermittency }
\author{Y. X. Huang\inst{1,2}\thanks{E-mail: \email{yongxianghuang@gmail.com}} \and F. G. Schmitt\inst{1}\thanks{E-mail: \email{francois.schmitt@univ-lille1.fr}}
\and Z. M. Lu\inst{2}
\and Y. L. Liu\inst{2}} \shortauthor{Y. X. Huang \etal}

\institute{
  \inst{1} Universit\'e des Sciences et Technologies de Lille - Lille 1, CNRS, Laboratory of Oceanology and Geosciences,
  UMR 8187 LOG,  62930 Wimereux, France\\
  \inst{2} Shanghai Institute of Applied Mathematics and Mechanics, Shanghai University, 200072 Shanghai,  China
}
\pacs{05.45.Tp}{Time series analysis}
\pacs{02.50.Fz}{Stochastic analysis}
\pacs{47.27.Gs}{Isotropic turbulence; homogeneous
turbulence}
\pacs{47.53.+n}{Fractals in fluid dynamics}

\abstract{ Hilbert-Huang transform is a method that has been
introduced recently to decompose nonlinear, nonstationary time
series  into a sum of different modes, each one having a
characteristic frequency. Here we show the first successful
application of this approach to homogeneous turbulence time series.
We associate each mode to dissipation, inertial range and integral
scales. We then generalize this approach in order to characterize
the scaling intermittency of turbulence in the inertial range, in an
amplitude-frequency space. The new method is first validated using
fractional Brownian motion simulations. We then obtain a 2D
amplitude-frequency representation of the pdf
 of turbulent fluctuations with a scaling trend, and we show how multifractal
exponents can be retrieved using this approach.  We also
find that the log-Poisson distribution fits the velocity amplitude
pdf better  than the lognormal distribution.  }

\begin{document}

\maketitle

\section{Introduction}
In nature and the real world, most data are nonlinear, nonstationary
and noisy, and general data-driven methods to analyze such data,
without \textit{a priori} assumptions basis, are demanded. About ten
years ago, such a method has been proposed to analyze nonlinear and
nonstationary time series: Hilbert-Huang transform (hereafter HHT)
\cite{huang1998emd,huang1999nvn}. The first step of this method is
the Empirical Mode Decomposition (EMD), which is used to decompose a
time series into a sum of different time series (modes), each one
having a characteristic frequency \cite{Wu2004a,flandrin2004emda}.
The modes are called  Intrinsic Mode Functions (IMFs) and satisfy
the following two conditions: (\romannumeral1) the difference
between the number of local extrema and the number of zero-crossings
must be zero or one; (\romannumeral2) the running mean value of the
envelope defined by the local maxima and the envelope defined by the
local minima is zero. Each IMF  has a characteristic scale which is
the mean distance between two successive maxima (or minima). The
procedure to decompose a signal into IMFs is the following:
\begin{itemize}
 \item[1]   The local extrema of the signal $X(t)$  are identified;
 \item[2]   The local maxima are connected together forming an upper envelope
 $e_{\max}(t)$, which is obtained by a cubic spline interpolation.
 The same is done for local minima, providing a lower envelope  $e_{\min}(t)$;
 \item[3]   The mean is defined as $m_1(t)=(e_{\max}(t)+e_{\min}(t))/2$;
 \item[4]   The mean is subtracted from the signal, providing the local detail
 $h_1(t)=X(t)-m_1(t)$;
 \item[5]   The component  $h_1(t)$  is then examined to check if it satisfies
 the conditions to be an IMF.  If yes, it is considered as the first IMF and denoted
 $C_1(t)=h_1(t)$. It is subtracted from the original signal and the first residual,
 $r_1(t)=X(t)-C_1(t)$  is taken as the new series in step 1.  On the other hand, if $h_1(t)$  is
 not an IMF, a procedure called ``sifting process'' is applied as many times as
 needed to obtain an IMF.  The sifting process is the following:  $h_1(t)$ is considered
 as the new data; the local extrema are estimated, lower and upper envelopes
 are formed and their mean is denoted  $m_{11}(t)$. This mean is subtracted
  from $h_1(t)$, providing  $h_{11}(t)=h_1(t)-m_{11}(t)$.
  Then it is checked again if $h_{11}(t)$  is an IMF. If not, the sifting process
  is repeated, until the component  $h_{1k}(t)$  satisfies the IMF conditions.
  Then the first IMF is  $C_1(t)=h_{1k}(t)$ and the residual $r_1(t)=X(t)-C_1(t)$
  is taken as the new series in step 1.
 \end{itemize}
The above sifting process should be stopped by a criterion
which is not discussed here: more details about the EMD  algorithm  can be found
 in refs. \cite{huang1998emd,huang1999nvn,flandrin2004emda,flandrin2004emdb,huang2003cle}.

After decomposition, the original signal $X(t)$ is  written as
a sum of IMF modes $C_{i}(t)$ and a residual $r_n(t)$
 \begin{equation}
X(t)=\sum_{i=1}^{N}{C_{i}(t)}+r_{n}(t)
\end{equation}
EMD is associated with Hilbert Spectral Analysis (HSA)
\cite{Cohen1995,Long1995,huang1998emd}, which is applied to each
mode as a time frequency analysis, in order to locally extract a
frequency and an amplitude. More precisely, each mode function
$C(t)$ is associated with its Hilbert transform $\tilde{C}$
\begin{equation}
\tilde{C}(t)=\frac{1}{\pi}
\int_{-\infty}^{+\infty}\frac{C(\tau)}{t-\tau}\, \upd \tau
\end{equation}
 and the combination of $C(t)$ and $\tilde{C}(t)$ gives the analytical signal
 $z=C+j\tilde{C}=\mathcal{A}(t)e^{j\theta(t)}$, where $\mathcal{A}(t)$
 is an amplitude time series and $\theta(t)$ is the phase of the mode
 oscillation \cite{Cohen1995}. Within such approach
and neglecting the residual, the original time series is rewritten as
\begin{equation}
X(t)=Re\sum_{i=1}^{N}{\mathcal{A}_{i}(t)}e^{j \theta_i(t) }
\end{equation}
where $\mathcal{A}_i$ and $\theta_i$ are the amplitude and phase time series of mode  $i$ and $Re$ means
real part \cite{huang1998emd,huang1999nvn}. For each mode, the Hilbert spectrum is defined as the square amplitude
 $H(\omega,t)=\mathcal{A}^2(\omega,t)$, where $\omega=d\theta/dt$ is the instantaneous frequency
extracted using the phase information
$\theta(t)=\tan^{-1}\tilde{C}(t)/C(t)$. $H(\omega,t)$ gives a local
representation of energy  in the time-frequency domain. The Hilbert
marginal spectrum of the original time series is then written as
$ h(\omega)=\int H(\omega,t) \, \upd t $ and corresponds
to an energy density at frequency $\omega$
\cite{Long1995,huang1998emd,huang1999nvn}.

Since its introduction, this method has attracted a large interest
\cite{huang2005emdbook}. It was shown to be an efficient method to
separate a signal into a trend and small scale fluctuations on a
dyadic bank
\cite{Wu2004a,flandrin2004emda,flandrin2004emdb}; it has
also been applied to many fields  including physiology
\cite{Su2008}, geophysics  \cite{Janosi2005}, climate studies
\cite{Sole2007}, mechanical engineering \cite{Chen2004} and
acoustics \cite{loutridis2005ril}, to quote a few.
 These studies showed the applicability of the so-called EMD-HSA approach on many different time series.
 In this letter, we apply the EMD and HSA approaches to fully developed turbulence time series. We first show
that the EMD method applies very nicely to turbulent velocity time series,  with an almost dyadic filter
bank in the inertial range. We then show how the HSA can be generalized to take into account intermittency.
We apply this to the turbulence time series, providing a first characterization of the intermittency of turbulence
in an amplitude-frequency representation.

%*****spectrum****Hilbert marginal spectrum and FFT of velocity*****
\begin{figure}[!htb]
\centering
\includegraphics[width=0.9\linewidth]{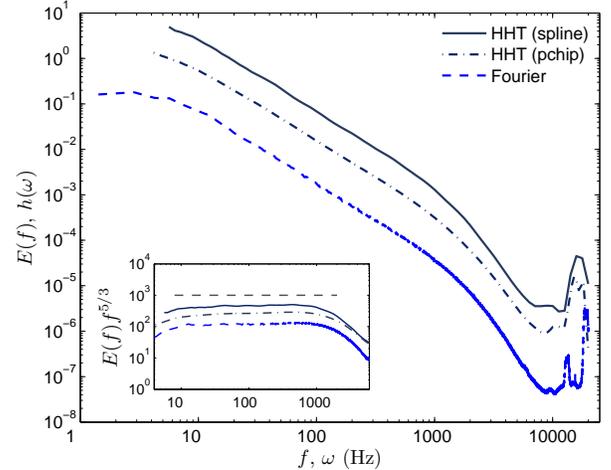}
\caption{Comparison of the Hilbert marginal energy spectrum (solid
line) and Fourier spectrum (dashed line, vertically shifted). The
slope of the  reference line is $-5/3$. Both the second order
Hilbert  and Fourier spectra indicate the same inertial subrange,
$10<f \,(\textrm{or }\omega)<1000 \, \textrm{Hz}$. The insert shows
the compensated spectra. The HHT spectra estimated using two different algorithms are shown for comparison, indicating a
 stability of the spectrum with respect to the algorithm used.
 }\label{fig:spectrum}
\end{figure}

\section{Application of EMD to turbulence time series}
We consider here a database obtained from measurements of nearly
isotropic turbulence downstream an active-grid
characterized by the Reynolds number $Re_{\lambda} =720$. The
sampling frequency
 is $f_{\mathrm{s}}=40 \un{kHz} $  \cite{kang2003dta}.
 The sampling time is $30\un{s}$ , and the total
number of data points per channel for each measurement is $1.2
\times 10^{6}$. We consider data in the streamwise direction at
position $x/M=48$, where $M$ is the mesh size and $x$ is the
distance in the streamwise direction. The mean velocity at this
location is $10.8 \un{ms^{-1}}$  and the turbulence intensity is
about $10\%$. For details about the experiment and the data see
ref.~\cite{kang2003dta}.

\begin{figure} \centering
  \includegraphics[width=0.9\linewidth]{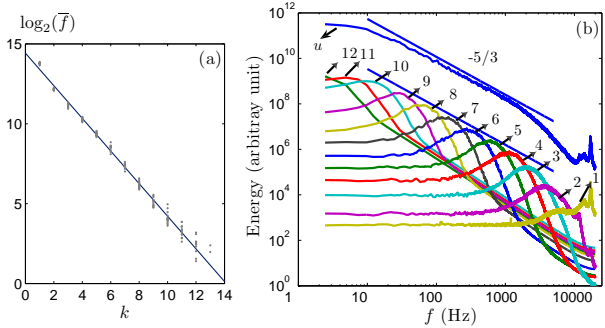}
% Here is how to import EPS art
\caption{(a) Mean frequency versus mode number for the turbulent
velocity time series.
  There is an exponential decrease with a slope very close to 1. This indicates that
  EMD acts as a filter bank which is almost dyadic. (b) Fourier spectrum of each mode  (from 1 to 12) showing that they
 are narrow-banded. The slope of the reference  line  is $-5/3$ corresponding to
  the inertial-range Kolmogorov spectrum.}\label{fig:scale}
\end{figure}

 Figure \ref{fig:spectrum} shows the second order Hilbert and Fourier spectra of the longitudinal  velocity.
 A Kolmogorov $-5/3$
 spectrum is observed in range $10 <f \,(\textrm{or }\omega)<1000 \un{Hz}$
for both spectra, indicating an inertial subrange over 2 decades.
Two different HHT spectra estimated using two different algorithms are shown in this figure:
 the very similar shape of the spectra indicates a
 stability of the spectrum with respect to the algorithm used.  The scaling which is obtained
  shows that Hilbert spectral analysis can be used to recover
Kolmogorov scaling in the inertial subrange.
The original velocity time series is divided into 73 non-overlapping segments
 of $2^{14}$ points each. After decomposition,
the original velocity series is decomposed into several IMFs from 11
to 13 modes with one residual. The time scale is increasing with the
mode; each mode has a different mean frequency, which is estimated
by considering the (energy weighted) mean frequency in the Fourier
power spectrum. The relation between mode number $k$ and mean
frequency~\cite{huang1998emd} is displayed in
fig.~\ref{fig:scale} (a). The straight line in log-linear
plot which is obtained suggests the following relation
$\overline{f}(k) = f_0 \rho^{-k} $, where $\overline{f}$
is the mean frequency, $f_0 \simeq 22000$ is a constant and $\rho =
1.9 \pm 0.1$ is very close to 2, the slight discrepancy
from 2 may be an effect of intermittency. This result may also slightly depend on the number of
iterations of the sifting process: in the present algorithm, the latter is variable but some
proposed algorithms contain a fixed maximum number of iterations.

This indicates that EMD
acts as a dyadic filter bank in the frequency domain; an analogous
property was obtained previously using stochastic simulations of
Gaussian noise and fractional Gaussian noise (fGn)
\cite{flandrin2004emda,flandrin2004emdb,Wu2004a}, and it is
interesting to note here that the same result holds for fully
developed turbulence time series, possessing long-range correlations
and intermittency \cite{Frisch1995}.

%\begin{figure} \centering
%  \includegraphics[width=0.9\linewidth]{pic/Fig3}
%% Here is how to import EPS art
%  \caption{Fourier spectrum of each mode  (from 1 to 12) showing that they
%  are narrow-banded. The slope of the reference  line  is $-5/3$ corresponding to
%  the inertial-range Kolmogorov spectrum.}\label{fig:Fourier}
%\end{figure}

We then interpret each mode according to its characteristic time
scale. When compared with the original Fourier spectrum of the
turbulent time series~(see fig.~\ref{fig:scale} (b)),
these modes can be termed as follows:
 the first mode, which  has the smallest time scale, corresponds to the measurement noise;
 modes 2 and 3 are associated with the dissipation range of turbulence.
 Mode 4  corresponds to the Kolmogorov scale, which is the scale below which
 dissipation becomes important; it is a transition scale between inertial range and
 dissipation range.
 Modes 5 to 10 all  belong to the inertial range corresponding
 to the scale-invariant Richardson-Kolmogorov energy cascade \cite{Frisch1995};
larger modes belong to the large forcing scales.
Figure~\ref{fig:scale} (b) represents the Fourier power
spectra of each mode. It shows that each mode in the inertial range
is narrow-banded. This confirms that the EMD approach can be used as
a filter bank for turbulence time series. In the next section, we
focus on the intermittency properties.

\section{Intermittency and multiscaling properties: Arbitrary order Hilbert spectral analysis}

Intermittency and multiscaling properties have been found in many
fields, including turbulence \cite{Frisch1995}, precipitations
\cite{Schertzer1987}, oceanography \cite{Seuront1999}, biology
\cite{Ashkenazy2002}, finance \cite{Schmitt1999}, etc.
 Multiscaling intermittency is often characterized using structure function of order  $q>0$
 as the statistical moment of the
fluctuations  $\Delta X_{\tau}=\vert X(t+\tau)-X(t)\vert$ (see ref.
\cite{Frisch1995} for reviews):
\begin{equation}
\langle (\Delta X_{\tau})^q \rangle \sim C_q \tau^{\zeta(q)}
\end{equation}
where $C_q$ is a constant and $\zeta(q)$  is a scale invariant moment function; it is also a cumulant generating function,
 which is nonlinear and concave and fully
characterizes the scale invariant properties of intermittency.

We present here a new
method to extract an analogous intermittency function using the EMD-HSA methodology.
 The Hilbert spectrum $H(\omega,t)$  represents the original signal at the local level.
 This can be used to
 define the joint probability density function (pdf) $p(\omega,\mathcal{A})$ of the frequency $[\omega_i]$
 and amplitude $[\mathcal{A}_i]$, which are extracted from  all modes $i=1\cdots N$ together. The Hilbert
marginal spectrum is then rewritten as
\begin{equation}
   h(\omega)=\int_0^{\infty} p(\omega,\mathcal{A}) \mathcal{A}^2 \, \upd \mathcal{A} \label{eq:marginal2}
\end{equation}
This definition corresponds to a second statistical moments. We then
naturally generalize eq.~(\ref{eq:marginal2}) into arbitrary
moments:
 \begin{equation}
 \mathcal{L}_q(\omega)=\int_0^{\infty} p(\omega,\mathcal{A}) \mathcal{A}^q \, \upd \mathcal{A}
 \label{eq:arbitrary}
\end{equation}
where $q \ge 0$ and $h(\omega)=\mathcal{L}_2(\omega)$
\cite{Huang2008TSI}. In the inertial range, we assume the
following scaling relation:
 \begin{equation}
 \mathcal{L}_q(\omega) \sim \omega^{-\xi(q)}
 \label{eq:arbitrary2}
\end{equation}
where $\xi(q)$ is the corresponding scaling exponent function in the
amplitude-frequency space. Equation~(\ref{eq:arbitrary}) provides
a new way to estimate the scaling exponents,
where, according to dimensional analysis, $\xi(q)-1$
can be compared  to $\zeta(q)$.

%******fBm2**Hilbert marginal spectrum of fBm****
\begin{figure}[htb]
  \centering
  \includegraphics[width=0.9\linewidth]{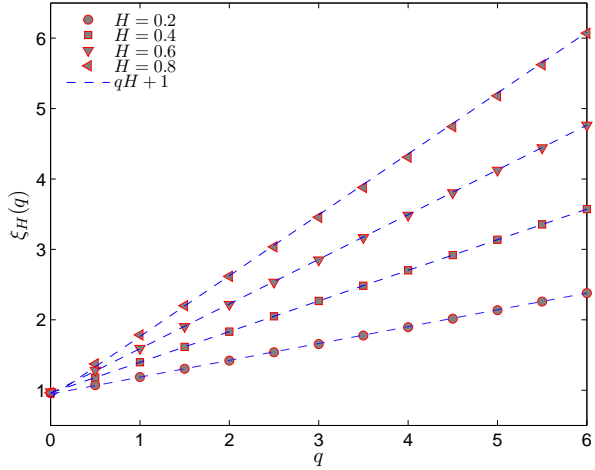}
  \caption{ Scaling exponents $\xi(q)$ for  fractional Brownian motion simulations with
     $H=0.2 $,  $0.4 $,  $0.6 $ and $0.8$, respectively. }\label{fig:fbm2}
\end{figure}

We first validate the new method by using  fractional Brownian
motion time series (fBm). They are characterized by the
Hurst number $0\le H \le 1$, and it is well-known that $\zeta(q)=qH$, hence we expect
$\xi(q)=1+qH$. We simulate 500 segments of length $2^{12}$ data
points each, using a wavelet based algorithm \cite{Abry1996}, with
different $H$ value
  from 0.2 to 0.8. The Hilbert transform is numerically estimated by using
a FFT based method \cite{marplejr1999cdt}. The scale invariance is
perfectly respected as expected, this is not shown here,
see ref.\cite{Huang2008TSI} for more detail on validations of the
method with fBm simulation.  We then represent the corresponding
scaling exponents $\xi(q)$ for various value of $q$ from 0 to 6, for
four values of
  $H$ ($H=0.2$, $0.4$, $0.6$ and $0.8$) in fig.~\ref{fig:fbm2}. The perfect straight lines of equation
$1+q H$ confirm the usefulness of the new method to estimate
$\xi(q)$.

%************jointpdf**2D representation of the pdf******************
\begin{figure}[htb]
\centering
\includegraphics[width=0.9\linewidth]{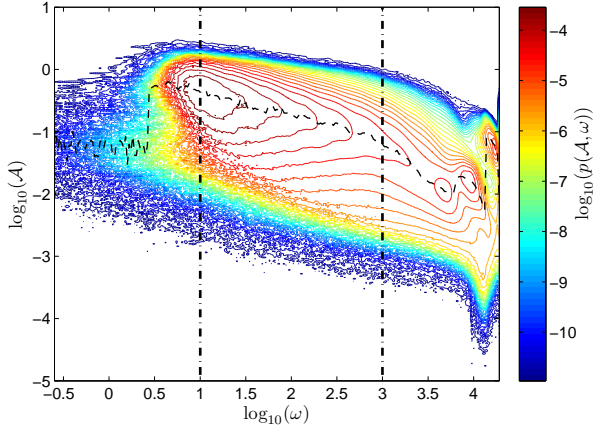}
\caption{Representation of the joint pdf ${p}(\omega,\mathcal{A})$
(in log scale) of turbulent fluctuations in an amplitude-frequency
space. The scaling range $10<\omega<1000 \un{Hz}$  for frequencies
is shown as vertical dotted lines. The dashed line shows the
skeleton $\mathcal{A}_{\mathrm{s}}(\omega)$ of the joint pdf, which
is the amplitude for which the  conditional pdf $p(\mathcal{A}\vert
\omega)$ is maximum.}\label{fig:jointpdf01}
\end{figure}

\begin{figure}[htb]
\centering
\includegraphics[width=.9\linewidth]{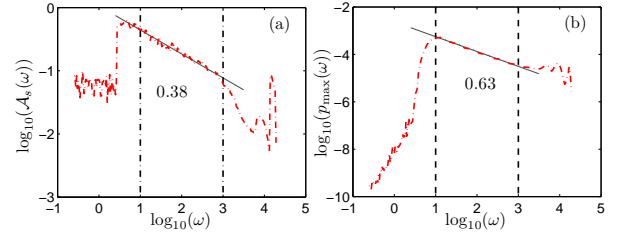}
\caption{The skeleton of the joint pdf. (a)
$\mathcal{A}_{\mathrm{s}}(\omega)$
 in log-log plot. A power law behaviour is observed in the
inertial subrange with scaling exponent 0.38, which is close to the
Kolmogorov value 1/3. (b) $p_{\max}(\omega)$
 in log-log plot.   A power law behaviour is observed
in the inertial subrange with scaling exponent 0.63. The vertical
lines show the corresponding inertial subrange
$10<\omega<1000\un{Hz}$. }\label{fig:jointpdf02}
\end{figure}

 \begin{figure}[htb]
\centering
\includegraphics[width=0.9\linewidth]{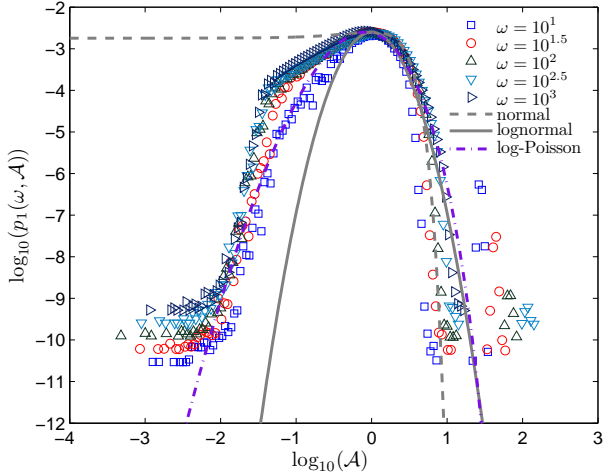}
\caption{Representation of the rescaled conditional pdf
$p_1(\mathcal{A},\omega)$ in the inertial range, for fixed values of
$\omega=10$, $10^{1.5}$, $10^{2}$, $10^{2.5}$ and $10^{3}
\un{Hz}$.}\label{fig:rescale}
\end{figure}

%************arbitrary**first eight positive integers order HMAS******************
\begin{figure*}[htb]
\centering
\includegraphics[width=0.7\linewidth]{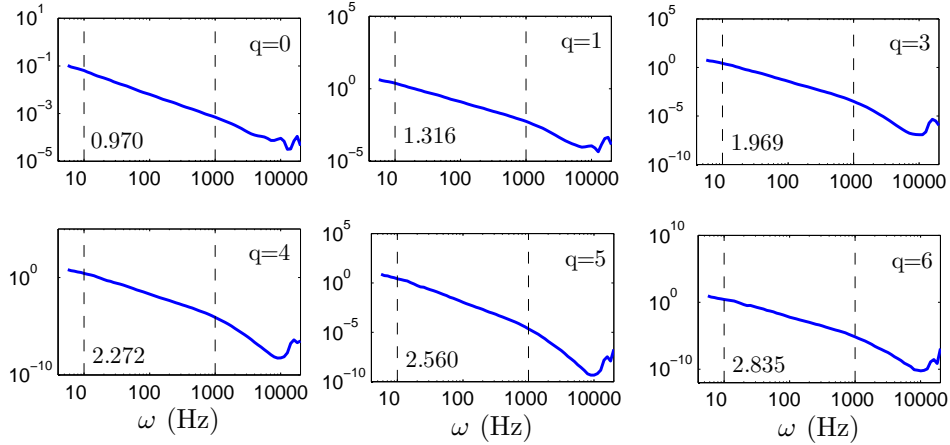}
\caption{Representation of $\mathcal{L}_{q}(\omega)$, Hilbert
spectral analysis of velocity intermittency, using different orders
of moments (0, 1, 3, 4, 5 and 6). Power laws are observed on the
range $10<\omega<1000 \un{Hz}$ for all spectra.  The value of the
scaling exponent $\xi(q)$ is shown in each
figure.}\label{fig:arbitrary}
\end{figure*}

%%************Scaling********************
\begin{figure}[htb]
\centering
\includegraphics[width=0.9\linewidth]{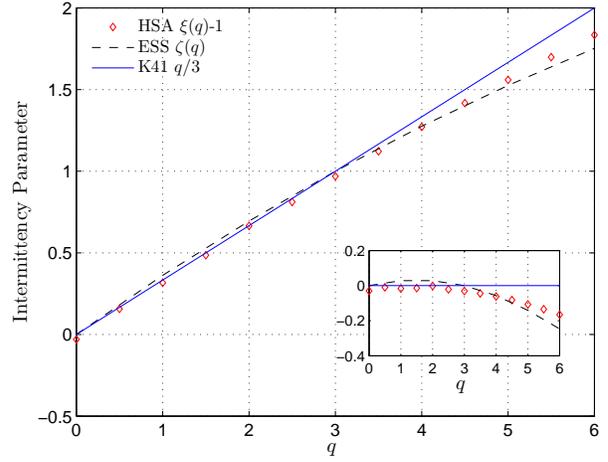}
\caption{Comparison of the scaling exponents   $\xi(q)-1$ (diamond)
with the classical $\zeta(q)$ obtained from structure functions
analysis with the ESS method (dash-dotted line) and K41 $q/3$ (solid
line). The insert shows the departure from the K41
law.}\label{fig:scaling}
\end{figure}

 We then consider turbulence intermittency properties using this approach.
The EMD-HSA methodological framework provides a way to represent
turbulent fluctuations in an amplitude-frequency space: the joint
pdf ${p}(\omega,\mathcal{A})$ is shown in fig.~\ref{fig:jointpdf01}.
The inertial  subrange for frequencies is shown as vertical dotted
lines. This figure is the first 2D amplitude-frequency
representation of the pdf of turbulent fluctuations; it can be seen
graphically that the amplitudes decrease with increasing
frequencies, with a scaling trend. We show in the same graph
 the skeleton $\mathcal{A}_{\mathrm{s}}(\omega)$ of the
joint pdf which corresponds to the amplitude for which the conditional pdf
$p(\mathcal{A}\vert \omega)$ is maximum:
\begin{equation}
  \mathcal{A}_{\mathrm{s}}(\omega)=\mathcal{A}_0\, ;\,
  p(\mathcal{A}_0,\omega)=\max_{\mathcal{A}}\{ p(\mathcal{A}\vert \omega)\}
  \end{equation}
We then reproduce the skeleton in fig.~\ref{fig:jointpdf02} in two
different views: (a) $\mathcal{A}_{\mathrm{s}}(\omega)$ in log-log
plot; (b) skeleton pdf $p_{\max}(\omega)=p(
\mathcal{A}_{\mathrm{s}}(\omega),\omega)= \max_{\mathcal{A}}\{
p(\mathcal{A}\vert \omega)\}$
 in log-log plot.
It is interesting to note that a power law behaviour is found for
both representations
\begin{equation}
  \mathcal{A}_{\mathrm{s}}(\omega)  \sim \omega^{-\beta_1},\,\,
  p_{\max}(\omega) \sim \omega ^{-\beta_2}
\end{equation}
where $\beta_1  \simeq 0.38$, and $\beta_2 \simeq 0.63$. Dimensional
analysis provides the non-intermittent Kolmogorov value
$\beta_1=1/3$ and $\beta_2=2/3$. The difference with these
theoretical value  may be an effect of intermittency. We note that
the value $\beta_1=0.38$ is comparable with the estimation of
$\zeta(1)=0.37$ given by Ref. \cite{Water1999}. We plot in
fig.~\ref{fig:rescale} the rescaled pdf $p_1( \mathcal{A},\omega) =
\omega^{\beta_2} p( \mathcal{A}/ \omega^{\beta_1},\omega)$, for
various fixed values of $\omega$. In case of monoscaling, these pdfs
should superpose perfectly; here the plot is scattered,
but nevertheless we note that   the lack of superposition of these
rescaled pdfs is a signature of intermittency. Moments of
this pdf are less noisy as will be visible bellow. For comparison,
we plot the normal distribution (dashed line), lognormal
distribution (solid line) and log-Poisson distribution
(dashed-dotted line) in the same figure. It seems that the
log-Poisson distribution provides a better fit to the pdf than the
lognormal distribution. We also characterize intermittency in the
frequency space by considering marginal moments
$\mathcal{L}_{q}(\omega)$. Figure~\ref{fig:arbitrary} shows
$\mathcal{L}_{q}(\omega)$, Hilbert spectral analysis of velocity
intermittency, using different orders of moments (0, 1, 3,
4, 5 and 6). The moment of order 0 is  the marginal pdf of the
instantaneous frequency, see eq.~(\ref{eq:arbitrary}). It is
interesting to note that this pdf is extremely ``wild'', having a
behaviour close to $\mathcal{L}_0(\omega)\sim \omega^{-1}$,
corresponding to a ``sporadic'' process whose probability density is
not normalizable
 ($\int p(\omega)\,\upd \omega$ diverges). This result is only obtained when all modes are considered together;
 such pdf is not found for the frequency pdf of an individual mode.
 This property seems to be rather general: we observed such pdf for moment of order zero using
 several other time series:
 for example surf-zone turbulence data, fBm \cite{Huang2008TSI}, river flow discharge data.
 Hence it does not seems to be linked to turbulence itself,
 but to be a main property of the HSA method, which still needs to be studied further.
 We observe the power laws in range $10 <\omega<1000 \un{Hz}$  for all order moments.
 The values of scaling exponents $\xi(q)$ are shown in each picture.
 This provides a way to estimate scaling exponents  $\xi(q)$ for every order of moment $q\ge
 0$ on a continuous range of scales in the frequency space.

Next, we compare scaling exponents $\xi(q)-1$ estimated by our new
approach with the classical structure functions scaling exponent
function $\zeta(q)$ estimated using the extended self similarity
(ESS) method \cite{Benzi1993} in fig.~\ref{fig:scaling}. It can be
seen that $\xi(q)-1$ is nonlinear and
  is  close to $\zeta(q)$, but the departure from K41 law shows that the curvature is not the same:
  $\xi(q)$ seems less concave than $\zeta(q)$.

  Here, we provide some comments on some issues of the EMD
  method. The main drawback of the EMD method is a lack of solid
  theoretical ground, since it is almost empirical \cite{huang2005emdbook}. It has been
  found experimentally that the method,  especially for the HSA, is statistically stable
  with different stopping criteria  \cite{huang2003cle}.  Recently,
  Flandrin \textit{et al.}  have obtained new theoretical results on
  the EMD method \cite{flandrin2004emdb,rilling2006ise,rilling2008oot}.
  However, more theoretical work is still needed to fully understand
  this method.

\section{Conclusion}

We have applied here empirical mode decomposition to analyze a
high Reynolds number  turbulent experimental
time series. After decomposition, the original velocity time  series
is separated into several intrinsic modes. We showed that this method acts as an
almost dyadic filter bank in the frequency domain, confirming
previous results that have been obtained on Gaussian noise or fractional
Gaussian noise. Comparing
the Fourier spectrum of each mode, and the associated characteristic scale,
we can interpret each mode according to the range to which it belongs.
The first mode contains the smallest scale and the  measurement noise;
two modes are associated to dissipation scales, and many modes are associated to
the inertial subrange corresponding to the turbulent energy cascade.
The last modes correspond to the large scales
associated to the coherent structures (energy-containing structures).

We have obtained a first 2D representation of the joint pdf
$p(\omega,\mathcal{A})$.    We observed a interesting power law
behaviour with scaling exponent $\beta_1\simeq 0.38$ for the
location  of the joint pdf skeleton points. We also observed a power
law behaviour with scaling exponent $\beta_2 \simeq 0.63$ for the
skeleton pdf $p_{\max}(\omega)$. It is also found that the
log-Poisson distribution provides a better fit to the velocity pdf
than the lognormal distribution. Then the intermittency information
in multiscaling (multifractal) turbulent processes was extracted
using the HSA framework.
 The scaling exponents in amplitude-frequency space
($\xi(q)-1$) are close to the ones in real space $\zeta(q)$, despite
the quite different approaches used in both cases.

We have here
extended the EMD-HSA approach in a quite natural way in order to
consider intermittency. This provides  a new time-frequency analysis
for multifractal time series, that is likely to be applicable to
other fields within the multifractal framework.

\acknowledgments This work is supported in part  by the National
Natural Science Foundation of China (No.10672096 and No.10772110)
and the Innovation Foundation of Shanghai University. Y.~H. is
financed in part by a Ph.D. grant from the French Ministry of
Foreign Affairs.  The EMD Matlab codes used in this paper are
written by P. Flandrin from laboratoire de Physique, CNRS \& ENS
Lyon (France): http://perso.ens-lyon.fr/patrick.flandrin/emd.html.
Experimental data have been measured in the Johns Hopkins
University's Corrsin wind tunnel and are available for download at
C. Meneveau's web page: http://www.me.jhu.edu/\~{}me\-neveau/datasets.html.
We thank the reviewers for helpful comments;

%\clearpage
 \bibliographystyle{eplbib}

\end{document}